# A COOPERATIVE ENTERPRISE AGENT BASED CONTROL ARCHITECTURE

Simona Caramihai, Ioan Dumitrache, Aurelian Stanescu, Janetta Culita

*Politehnica University of Bucharest, Dept. of Automatic Control and Computer Science*
*E-mail(s): sic@ics.pub.ro , idumitrache@ics.pub.ro, ams@utis.cpru.pub.ro,*
*jculita@yahoo.com*

Abstract: The paper proposes a hierarchical, agent-based, DES supported, distributed architecture for networked organization control. Taking into account enterprise integration engineering frameworks and business process management techniques, the paper intends to apply control engineering approaches for solving some problems of coordinating networked organizations, such as performance evaluation and optimization of workflows.

## 1. INTRODUCTION

The globalization of business and commerce makes enterprises increasingly dependent on their cooperation partners, clients and suppliers. At present, competition occurs between networks of enterprises instead of between individual enterprises. In this competition, the capabilities for interoperability of the enterprises become critical.

Interoperability has been used initially for the analysis of software systems, from the process control point of view. This means that a software system is actually composed by a number of subsystems able to communicate with each other and with the environment through well defined interfaces. All the communications are based on pre-defined function implementation.

The most important challenge raised by the new interpretation of the interoperability idea is how to represent the system specification semantics according to the given purpose. The recent research has highlighted the following conclusions:
- Interoperability covers technical, semantic and pragmatic interoperability (Berre, 2005).
- Technical interoperability means that messages can be transmitted from one application to another.
- Semantic interoperability means that the message content becomes understood in the same way by the senders and the receivers. This may require transformations of information representation or messaging sequences.
- The pragmatic interoperability captures the willingness of partners for the actions necessary for the collaboration. The willingness of participating presumes both capability of performing a requested action and policies dictating whether the potential action is preferable for the enterprise to be involved in.

Based on these premises, recent publications have revealed that the interoperability only at data and information level is not longer sufficient. **The collaboration must be done at knowledge level.** At this level all the organizational interoperability senses can be applied. For an efficient knowledge sharing process, a new research field appeared: **knowledge management (KM).**

Knowledge and particularly the capacity to manage, create and share knowledge, are becoming now the centre of the scope for a successful company. Knowledge becomes another attribute of the changing economic reality. KM and the learning culture of the company, as a "necessary" condition for Knowledge Economy (KE) represent an attitude and a way of working with management. This process is one of redefining the target of the company from a profit making or share value increasing entity to a knowledge creating and sharing unit.

In order to build a learning organization or a corporate learning culture, companies should be skilled in systematic problem solving, learning from their own experience and from the others, processing knowledge quickly and efficiently through the organization and experimenting with new approaches.

There is a certain necessity to create a new type of company with high autonomy, by using a distributed intelligent system with most important attributes: perception, learning, thinking, communication and planning. The large FMS (Flexible Manufacturing Systems) could be viewed as an Adaptive Complex Dynamic System with a high capacity to be reconfigurable and a large knowledge base. The feed forward and feedback control strategies are integrated on the general philosophy of knowledge management into a company.

The process of KM means the identification and analysis of knowledge, for developing new knowledge that will be used to realize organizational goals. Knowledge is gathered from a geographical and informational distributed system.

The best choice to define and to model a distributed system is represented by **multi-agent systems (MAS)**. A MAS is defined by modularity, abstraction and intelligence. An intelligent agent must be autonomous, but in the same time it has to develop the capacity of interoperability with the other entities of the system. The concept of autonomy can be interpreted in different ways (Wooldrige, 2002) by **reactivity, pro-activity and socialization.**

Usually, the agent strategy consists in an ordered set of activities and constraints – and can be modeled by **workflows**. Knowledge management implies, in this case, the appropriate choice of knowledge that can be used for partial or full completion of the workflow tasks. Moreover, a new workflow can be designed using pieces of other workflows.

Agent's interoperability must be translated into identification of the knowledge that can be used for the accomplishment of each other goals. This knowledge, having a structured modularity and representation, can be the basics for the collaboration support between the agents that have different modeling technologies and also that are physically placed into another structure. After the classes of knowledge have been identified, the simulation of business objective workflows can be done in order to determine the failure and success rate. Different applications (manufacturing systems, business processes and also information technology (Brooke, *et al.*, 2003), make obvious that there is necessary this type of approach.

Another challenge is related to the existence of a support for diagnostics. Few interoperability architectures support simulation, verification and validation of process designs. Few systems support collection and interpretation of real-time data. (van der Aalst, *et al.*, 2003). Anytime an agent can decide that it no longer desires to share its knowledge, therefore it is necessary to define protocols and standards for the recovering – simulation becomes critical for finding the scenarios when this approach is useful.

The paper intends to present an original approach for solving the above-mentioned problems in terms of control engineering – architecture, modelling support and evaluation. This approach is based on intelligent agent-based control architecture for Flexible Manufacturing Systems (FMS), derived itself from agent-based negotiation procedures proposed by Lin and Solberg (1994) and from holonic architectures (Wyns, *et al.*, 1999). It was based on a Petri Nets (PN) Discrete Event System modeling framework, allowing control policies synthesis and evaluation.

The second part of the paper will shortly introduce control concepts and the above mentioned architecture. The third one will present its extension for networked organisation as well as an illustrative case study. Conclusions will include some comments and further research directions.

## 2. THE SUPERVISED CONTROL ARCHITECTURE

The control architecture (Caramihai, *et al.*, 2001) was designed to have two main levels: the agent level and supervisory one.

The agent level has extended autonomy and communication capabilities. If the system is not perturbed, the agents will act following given rules and according to some pre-established patterns of association, with a global behavior similar to a hierarchical architecture. When a perturbation occurs, agents have to take their own decisions, acting as autonomous units. The agent-level itself acts as a real-time control level for the manufacturing system. On the other hand, the supervisory level has a more complex behavior: *on-line functioning* for monitoring the agents and taking some decisions on their performances and an *off-line supervisory[1] policy synthesis* for ensuring robustness and efficiency in the FMS global control.

The generic structure of an agent includes the following basic blocks:
- **The World Model block** - is a knowledge base containing the internal model of the agent (including relevant knowledge about the real world, i.e. the state of temporary links with other agents, the state of the controlled processes, the long and short-term goals of the agent and the value of the criteria of their achievement) and a rule base on "how to take decisions". The world model must have intrinsic updating mechanisms for its two main parts, verifying and ensuring the global consistency of data;
- **The decision making block** – it decides either how to process the received data (with direct and explicit consequences on the immediate actions of the control agent) or how to modify information and knowledge contained in the world model block (with implicit consequences on the whole behavior of the agent);
- **The perception block** - receives information from the environment, *via* the **Sensors** block and modifies, if necessary, the internal model of the agent;
- **The actuators block** - is basically an interface with the environment – **the real world**: in this situation it is rather a communication interface, either with the control module of the controlled process or with the other agents
- **The sensors block** – is the interface between the real world and the perception block, usually consisting of sensors and transducers with the purpose to record modifications induced by actuators on the real world. In this case, it is also a communication interface with the control module of the controlled process (for updates in the world model block) and with other agents.

The generic agent structure could be conceived as being "split" in two components:
- The **computational agent** part is working with symbolic information and has as modules the perception, world model and decision making blocks;
- The **real agent** part is composed of the sensors and the actuators blocks and works with the environment (the real world), modifying it and recording the effect of both its own actions and of those of other agents.

---

[1] Supervisor – the term is used with its both connotations: 1. supervisory level, monitoring and, if necessary, correcting agent activities, and 2. supervisor in the sense given by the classical DES theory; when using the "synthesis of the supervisor" expression, the second connotation is taken into account.

The two agent components act as a closed loop structure, where the real agent is the controlled process and the computational one is an adaptive control module. It is important to note that the actions of other agents are considered as perturbations in the real world, for which, because of agent-based philosophy, the world model is only a partial model.

It was assumed that a FMS can be modeled exclusively by two basic types of agents, each one having a specific internal organization: the *product agent* and the *resource agent*. The desired functioning of the FMS is achieved by the negotiations between the different types of agents.

A **product agent** is in charge with the completion of a specific product or activity. In order to accomplish its objective it negotiates with resource agents the necessary operations and tasks, following a given workflow model. A product agent owns a "manufacturing account" consisting in virtual cost units, which it can use for "paying" every operation it requires from other agents. Its goal is to finalize the products or activities it is responsible with following the associated workflow, in a given time interval, with minimum manufacturing cost and an adequate quality.

A **resource agent** controls a resource in order to ensure the required processing for product agents. It is supposed that it acts as a high level control structure and communicates with the real-time control module of the resource for obtaining information about its state and for starting specific tasks. Its goal is to maximize the amount of cost units received in a given time interval.

The internal model of a resource agent should contain at least: the list of processing tasks that the resource could perform - with their respective duration and cost - and the current state of the resource.

The agents can communicate, creating an interconnected structure. One of the main implementation problems of an agent-based architecture is derived from the large amount of inter-agents communication, which could become difficult to manage and overload the software system. The architecture presented in this paper restricts this communication flow by the fact that an agent has to contact only agents of different type.

The PN part of the world model of a resource agent takes into consideration only the two aspects mentioned above. The detailed internal model of the agents is not needed because the control architecture behavior is based on the global model. The PN model of a product agent describes the workflow it has to implement. The global model of a processing FMS is obtained by the synchronous composition of the respective resource and products PN models.

The supervisor presents two functioning regimes: one for the off-line synthesis and one for the on-line behavior.

When the system is working **off-line**, the supervisor have to be determined by firstly qualitatively analyzing the PN model of the FMS for ensuring that the system functioning will respect the desired specifications, without deadlocks or forbidden states. After performing the state-space analysis, the synthesis process determines a desired functioning (based on optimality criteria, as the time spent in the system by a product agent) and, based on this, it will prescribe an "optimum functioning" for agents. This prescribed functioning is represented as recommended partners for negotiations.

During its **on-line** regime, the supervisor monitors the system. The monitoring is aimed to verify that the agents respect their recommended evolutions. If a perturbation occurs, then the global system will start a degraded functioning regime with respect to the pre-established optimal one.

In this paper the breakdown of a resource has been considered as perturbation. Obviously, the respective resource agent will no longer participate to the negotiations. This will degrade the functioning of the product agents whose recommended partner it was. Furthermore, if the break-down has occurred during a processing task, then a product agent could be more specifically affected and the supervisor will have to take some decisions concerning it.

Other kind of decisions the supervisory module should take during the on-line functioning is linked especially with agent-based architectures and manufacturing processes specificities. Other kind of decisions is concerned with the proper resource sharing and with eventual bottlenecks and undesired states that can appear in the degraded functioning regime. The supervisor is still in possession of the overall model of the system and can simulate on it the results of achieved negotiations in order to estimate in advance the problems that can occur.

A procedure for the supervisor synthesis was presented in detail in (Bratosin, *et al.*, 2005) and there was designed a software application implementing it. The synthesis methodology is highly general and it can be applied for any size of the system.

## 3. ENTREPRISE CONTROL ARCHITECTURE

At enterprise level, the main problem is how to implement a general control architecture allowing to different and heterogeneous structures to share knowledge without having a predetermined idea on the actual goal and without having to modify the communication interfaces.

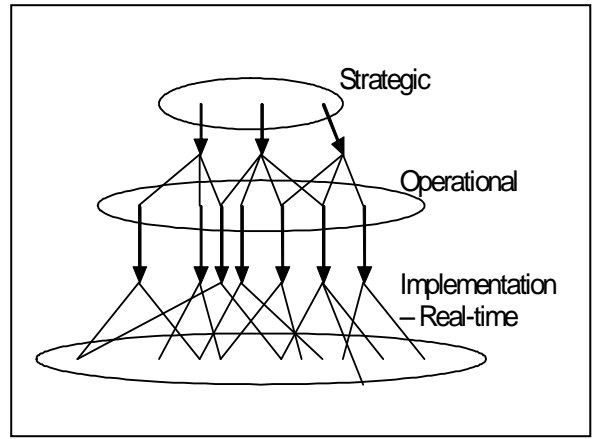

Fig. I Distributed control architecture

The main idea of the design is to realize a distributed intelligent architecture, where highly general activities initiated at the strategic level should be stepwise decomposed in more particular activities, using partial workflow models. Figure I illustrates the composition of partial workflow models into complex models. Basically, every workflow unit should encompass a "piece of knowledge" used for solving a given type of problem. So, the global knowledge of the system is distributed, and every agent acting with respect to a given workflow will include a part of the global intelligence of the system. This characteristic will result only by the interaction of agents, without requirements of real intelligence at agent level. Moreover, the actual management and evaluation of knowledge are performed at workflow level, and they do not depend on the actual implementation. At the strategic level, there are processes to be defined, each of them implying the planning and execution of a sequence of activities.

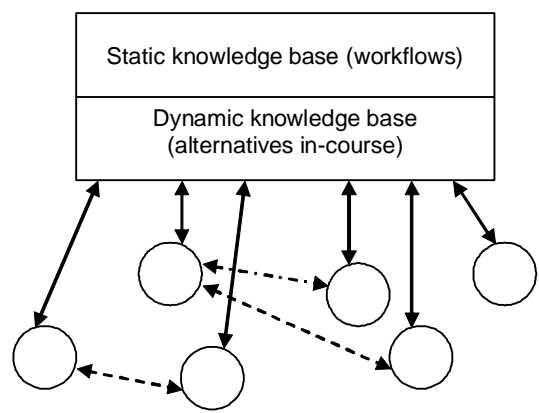

Fig. II Knowledge bases - agents interfaces

Agents could have two different types of communications: with the knowledge base (dynamical part) – *via* a dedicated interface, and with some of the other agents, for eventual negotiation and cooperative tasks. Static and dynamic knowledge-bases are in a general format representation (as Petri Nets, for instance).

The actual implementation of agents will necessitate communication interfaces with the dynamical knowledge base, but the knowledge transfer from one agent to another could be made at the workflow level, as well as the combination of different workflows in order to design new activities (Figure II). From the management level point of view, the static part of the database can be the support for strategic decisions and conception and the dynamic part for the operational ones.

The structure provides great flexibility in the choice of workflow alternatives and possibilities of dynamic reconfiguration in failure situations, in the sense described in Section 2. Moreover, it can be extended from shop-floor level until to networks of enterprises by the basic extension of the knowledge base, and without regard to the actual implementation of control agents; as well as the workflow definition is performed and the interfaces between the workflow representation and real-time control part exist.

## 4. CASE STUDY

The following case-study aims to illustrate how workflows can be modeled by Petri Nets and, consequently, different pieces of workflows can be combined to obtain a new one. It will be underlined how this composition can affect the execution of initial workflows and how limited resources sharing will result in several possibilities of activity scheduling, therefore necessitating a decision support system for choosing the most appropriate one.

Let be three different activities performed by different actors that have to share common resources. At operational level they can be executed with the following constraints:
- A1 consists of three jobs: J11, J12, J13, that are executed strictly sequential;
- A2 consists of jobs J21, J22, J23, where J21 should be performed first; the other two can be performed in any order, but sequentially, as long as they are starting after J21.
- A3 consists of jobs J31, J32, J33, where J33 is the last to be performed, staring after the end of the other two, which can be performed in any order, including concurrency.

Each job needs some resources, as presented in Table 1. There are two resources R1 available and one of each other type. The workflows representing the activities are contained at the static knowledge base level while the global model including them together with their relationships at the dynamical knowledge base level. The global model could be stored for an eventually employment, after that being removed from its location.

Table 1 Resource allocation and time duration for each component job of workflow activities

| Job | J11 | J12 | J13 | J21 | J22 |
|---|---|---|---|---|---|
| Res. (time) | R1(3) R3(3) | R3(1) | R4(2) | R1(2) | R2(1) |

| Job | J23 | J31 | J32 | J33 |
|---|---|---|---|---|
| Res. (time) | R3(2) | R1(3) | R4(2) | R2(2) |

Figure III illustrates the global workflow model for this case study.

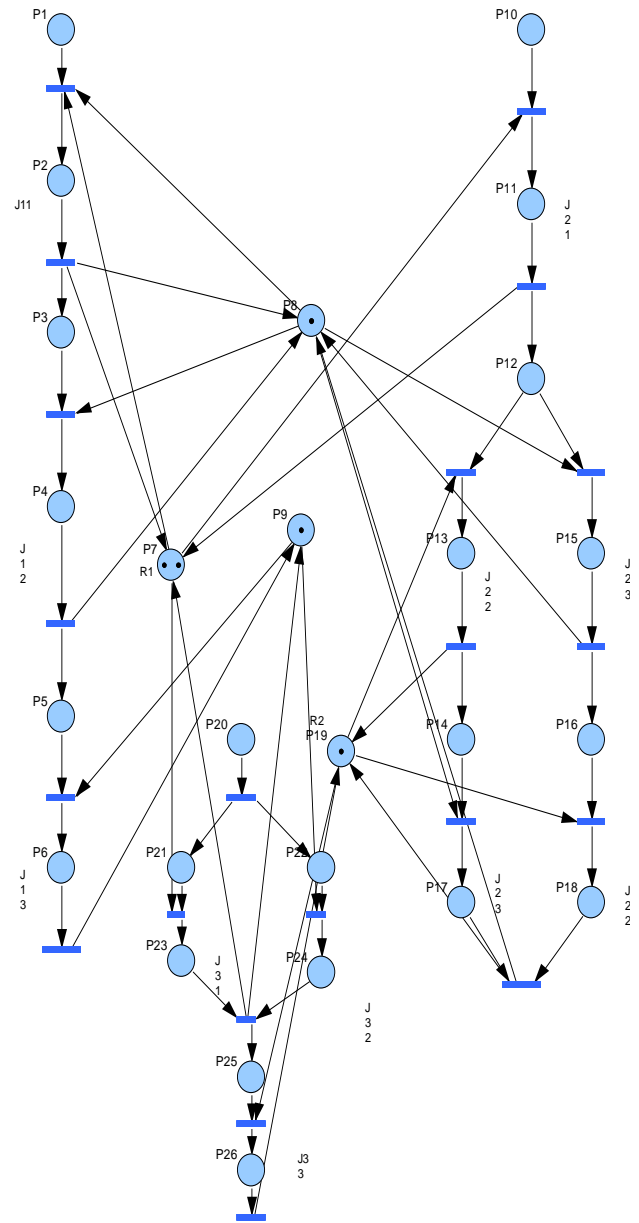

Fig. III Petri net global model

From the modeling point of view, there are some representation rules for workflows:
- workflows in static database are only represented at a structural level, thus representing jobs, job sequencing with execution constraints and necessary resources;

- real-time process data as number of entities to be processed (markings of P1, P10, P20) according to workflow and actual resource capacities (marking of P7, P8, P9, P19) – i.e. the complete initial marking of the Petri Net are specified (by an initialization procedure) only in the dynamic part of the data base;
- every job is represented by a block transition-place-transition, as its execution is considered indivisible (places meaning "job in execution" and transitions "start" and respectively "end" job);
- between every two successive jobs, there is a place with the meaning of "waiting", in order to avoid deadlocks due to resources sharing.

It could be observed that even for a unary initial marking of P1, P10, P20, there are several effective conflicts that will affect the execution of activities, thus delaying some of them.

Using the software application mentioned in Section 2 (Caramihai, *et al.*, 2005), the above mentioned initial marking will result in 8 execution possibilities that the application will underline. Therefore, allocating an adequate cost function for every activity, priorities can be established in order to obtain the "better" activity scheduling for the context.

## 5. CONCLUSION

The paper presents an architectural approach, based on agent-based knowledge management, Petri Net modeling and interoperability for cooperative enterprise networks.

The approach continues the previous work in supervisory synthesis for agent-based architectures and it offers the foundation for a Decision Support System in activity planning.

## REFERENCES

Berre, A-J (2005). "State-of-the art for Interoperability architecture approaches" with a focus on "Model driven and dynamic, federated enterprise interoperability architectures and interoperability for non-functional aspects" Retrieved June 13, 2006, from http://interop-noe.org/deliv/DAP1B.

Bratosin, C, S. Caramihai and H. Alla (2005). Supervisory Synthesis for Safe Petri Nets. In: *Proceedings of the 15th International Conference on Control Sytems and Computer Science* (I.Dumitrache & C.Buiu (Eds)), Vol. 1, pp. 149-154. Politehnica Press, Bucharest, RO.

Brooke, J., K. Garwood and C. Goble (2003). Interoperability of Grid Resource Descriptions: A Semantic Approach. Retrieved May 15, 2004, from http://www.semanticgrid.org/GGF/ggf9/

Caramihai, S., C. Bratosin and C. Munteanu (2005). A DES – supported agent based control architecture for FMS. In: *17th IMACS World Congress, CD-ROM,* Paris, France.

Caramihai, S., C. Munteanu, J. Culita and A. Stănescu (2001). Control Architecture for FMSs based on Supervised Control of Autonomous Agent. In: *Proceedings of the 13th International Conference on Control Systems and Computer Science* (I.Dumitrache & C.Buiu (Eds)), Vol. 1, pp. 164-169. Politehnica Press, Bucharest, RO.

Lin, Y and Solberg, J (1994). Autonomous control for open manufacturing systems. In: *Computer Control of Flexible Manufacturing Systems*. (S. Joshi and J. Smith (Eds)), pp. 169-206. Chapman & Hall, London, UK.

Wyns, J., H. Van Brussel and P. Valckenaers (1999). Design pattern for deadlock handling in holonic manufacturing systems. In: *Production Planning and Control*, Vol. **10**, No. 7, pp. 616-626.

Van der Aalst, W. M. P, A. H. M. ter Hofstede and M. Weske (2003). Business process management: A survey. In: *Proceedings of the International Conference on Business Process Management* (W.M.P. van der Aalst, A.H.M. ter Hofstede and M. Weske (Eds)), Volume Lecture Notes in Computer Science, pp. 1-12. Springer-Verlag, Berlin.

Wooldridge, M. (2002), *An Introduction to Multi Agent Systems*, John Wiley & Sons Ltd, New York.